\documentstyle[12pt,epsfig,epsf]{article}

\newcommand{\be}{\begin{equation}}
\newcommand{\en}{\end{equation}}
\newcommand{\bfx}{\vec{r}}
\newcommand{\rhoi}{\rho_{\mathrm{i}}}
\newcommand{\rhom}{\rho_{\mathrm{m}}}
\newcommand{\ndef}{n_{\mathrm{def}}}
\newcommand{\psilok}{|\psi_{6,i}|^2}

\def\gsim{\hbox{\lower3pt\vbox{\baselineskip=4pt \lineskiplimit=0pt \kern2pt 
          \hbox{$>$}\hbox{$\sim$}}}}

\begin{document}

%\draft
                                                                              
\title{Computer simulations of the two-dimensional \\
melting transition using hard disks} 

\author{A.\ Jaster \\[4mm]
Universit\"{a}t - GH Siegen, D-57068 Siegen, Germany}
\date{\today}
                                                                                
%%%%%%%%%%%%%%%%%%%%%%%%%%%%%%%%%%%%%%%%%%%%%%%%%%%%%%%%%%%%%%%%%%%%%%%% 

\maketitle

\begin{abstract}
We present detailed Monte Carlo results for the 
two-dimensional melting transition of various systems
up to $N=65536$ hard disks.
The simulations are performed in the $NVT$ ensemble, using a
new updating scheme. In the isotropic phase 
the bond orientational correlation length $\xi_6$   and the
susceptibility $\chi_6$ are measured  
and compared  with the predictions of the 
Kosterlitz-Thouless-Halperin-Nelson-Young (KTHNY) theory.
From the scaling relation of $\xi_6$ and $\chi_6$ 
we calculate the critical exponent $\eta_6$.
In the phase transition region we use finite-size 
scaling methods to locate the disclination binding transition point
and compare the results with the values obtained from the behaviour in
the isotropic phase. Additionally, we measure 
the topological defect density, the pressure and
the distribution of the second moment of the 
local bond orientational order parameter.
All results are in good agreement with the KTHNY theory, while a 
first-order phase transition with small correlation length 
and a one-stage continuous 
transition can be ruled out. 
\\[4mm]
\end{abstract}

%\pacs{64.70.Dv, 64.60.Fr}

%%%%%%%%%%%%%%%%%%%%%%%%%%%%%%%%%%%%%%%%%%%%%%%%%%%%%%%%%%%%%%%%%%%%%%%%

%% \begin{multicols}{2} \narrowtext

%%%%%%%%%%%%%%%%%%%%%%%%%%%%%%%%%%%%%%%%%%%%%%%%%%%%%%%%%%%%%%%%%%%%%%%%   
%%%%%%%%%%%%%%%%%%%%%%%%%%%%%%%%%%%%%%%%%%%%%%%%%%%%%%%%%%%%%%%%%%%%%%%%%
\section{Introduction}        
The nature of the two-dimensional  melting transition 
is a long unsolved problem \cite{STRAND,GLACLA}. Melting
in two dimensions differs from the three-dimensional 
case because  the two-dimensional 
solid posses only quasi-long-range
positional order, while the three-dimensional solid is truly
long-range positional ordered. This means that the  
correlation function in two dimensions decays algebraically to 
zero for large distances, while it decays to a non-zero value
in three dimensions. This absence of conventional long-range
order at non-zero temperatures in two dimensions was pointed out by
Mermin and Wagner \cite{MERWAG}. Therefore, the mean-square displacement
of the particles from their ideal lattice position will diverge
logarithmically with the size of the system 
and no Bragg peaks in a strict manner
can occur in the thermodynamic limit.  Nevertheless, the
other order parameter, which describes the bond orientational 
order, is truly long-ranged ordered, i.e.\ the orientation of 
the bonds between neighboured particles is correlated over arbitrary
distances. 

There are several theoretical approaches for the description
of melting in two dimensions.
Halperin and Nelson as well as Young have developed a theory 
based on the the ideas of Kosterlitz and Thouless \cite{KTHNY}. 
The KTHNY theory deals with 
unbinding scenarios of topological defects, where
the two order parameters are related to two different topological defects:
the disclinations and the dislocations. 
The dislocation unbinding at a temperature $T_{\mathrm{m}}$
is responsible for the melting transition, while the
disclination unbinding at $T_{\mathrm{i}}$ 
destroys the bond orientation.
The first continuous transition transforms the solid into a 
novel hexatic phase,
which is short-ranged positional and quasi-long-ranged orientational
ordered. The second continuous  transition transforms this hexatic phase
in an isotropic one, i.e.\ a phase with short-ranged positional
and orientational order. 
An alternative scenario has been proposed by Chui \cite{CHUI}.
He presented a theory via spontaneous generation of grain boundaries, i.e.\ 
collective excitations of dislocations. He found that grain
boundaries may be generated before the dislocations unbind if the 
core energy of dislocations is sufficiently small, and predicted a 
first-order transition. This is characterized by a coexistence region of the
solid and isotropic phase, while no hexatic 
phase exists. Another proposal was given by Glaser and Clark \cite{GLACLA}.
They considered a detailed theory where the transition is handled as 
a condensation of localized, thermally generated geometrical defects 
and found also a first-order transition.
Calculations based on the density-functional approach were done
by Ryzhov and Tareyeva \cite{RYZTAR}. They derived that the hexatic phase 
cannot exist in the hard disk system.

Numerical investigations of two-dimensional melting can be done
in several ways. On the one hand one can simulate the particle system
or  the defect system \cite{SAITO}, on the other hand one can study
lattice models which describe
defects and their elastic interaction \cite{KLEINERT}
or grain boundaries \cite{JASHAH}. 

The hard disk system is one of the simplest particle models
to study the melting transition in two dimensions
with computer simulation techniques.
Even for this simple case no consensus about
the existence of an hexatic phase has been established.
The melting transition of the hard disk system was 
first seen in a computer simulation by Alder and Wainwright 
\cite{ALDWAI}. They used a system of $N=870$ disks, constant volume $V$ 
and molecular dynamics methods ($NVE$ ensemble)
and found that this system undergoes a first-order phase
transition from the solid to the isotropic phase. However,
the results of such small systems are affected by large finite-size 
effects. Simulations performed in the last years used 
Monte Carlo (MC) techniques either with constant volume 
($NVT$ ensemble) \cite{ZOCHLE,ZOLCHE,WEMABI,MIWEMA} or constant
pressure ($NpT$ ensemble) \cite{LEESTR,FEALST}.
Zollweg, Chester and Leung \cite{ZOCHLE} made detailed investigations
of large systems up to $16384$ particles, but draw no conclusives 
about the order of the phase transition.
The analysis of Zollweg and Chester \cite{ZOLCHE} for the pressure gave
an upper limit for a first-order phase transition, but is
compatible with all other scenarios. Lee and Strandburg \cite{LEESTR} 
used isobaric MC simulations and 
found a double-peaked structure in the volume distribution.
Lee-Kosterlitz scaling led them to conclude that the phase transition
is of first order. However, the data are not in the scaling region,
since their largest system contained only 400 particles.
MC investigations of the bond orientational order
parameter via finite-size scaling with the block analysis technique 
of 16384 particle systems were done by
Weber, Marx and Binder \cite{WEMABI}. They also 
favoured a first-order phase transition. 
In contrast to this, 
Fern\'{a}ndez, Alonso and Stankiewicz \cite{FEALST}\footnote{For
a critical discussion of this work see Ref.\ \cite{WMFAS}.} 
predicted a one-stage continuous melting
transition, i.e.\  a scenario with a single continuous transition
at $\rhoi=\rhom$
and consequently without an hexatic phase.
Their conclusions were based on the examination of the bond orientational 
order parameter in very long runs of different systems up to 15876
particles and hard-crystalline wall boundary conditions.
Finally, Mitus, Weber and Marx \cite{MIWEMA} studied 
the local structure of a system with $4096$ hard disks.
From the linear behaviour of a local order parameter they derived
bounds for a possible coexistence region.

In a recent letter \cite{JASTER2} we  published the first results
of simulations of the hard disk model 
in the $NVT$ ensemble with up to $65536$ particles 
to answer the question of the kind
of the phase transition. 
We showed that the behaviour of the susceptibility $\chi_6$ and
the bond orientational correlation length $\xi_6$ in the 
isotropic phase as well as the
value of the critical exponent $\eta_6$ coincide with the predictions
of the KTHNY theory. 
Additionally, we performed finite-size scaling (FSS) investigations
in the transition region and showed that these results are also in agreement
with the KTHNY scenario. Here we discuss the methods  
in detail and present additional results for the
pressure, the topological defect density and the distribution
of the second moment of the local bond orientational order parameter.
All results are compared with the predictions of the 
KTHNY theory.

%%%%%%%%%%%%%%%%%%%%%%%%%%%%%%%%%%%%%%%%%%%%%%%%%%%%%%%%%%%%%%%%%%%%%%%%   
%%%%%%%%%%%%%%%%%%%%%%%%%%%%%%%%%%%%%%%%%%%%%%%%%%%%%%%%%%%%%%%%%%%%%%%%%
\section{Algorithm and measurement}
As mentioned before, we used MC techniques and the $NVT$ ensemble
for the simulations of the hard disk system.
The updating was performed with an improved (non-local) Metropolis algorithm 
\cite{JASTER1}. 
We consider systems of $N=32^2$, $64^2$,
$128^2$ and $256^2$ hard disks in a two-dimensional square 
box. We find that finite-size effects with these
boundary conditions are 
not substantially larger than in a rectangular box with ratio 
$\sqrt{3}:2$ since no simulations in the solid phase were made.
This point will be discussed later.
The simulations
were performed on a Silicon Graphics workstation and a CRAY
T3E. The CPU time for the CRAY was of the order of some
month per node, where we have used 7 or 8 nodes. Further details are 
described in Ref.\ \cite{JASTER2}.

Careful attention has been paid to the equilibration of all systems.
We controlled that the expectation values had stabilized over long time.
Additionally, we measured some autocorrelation times for smaller
systems \cite{JASTER1} and estimated the values of larger systems
(for large correlations lengths) by assuming $z \approx 2$. 
We spent at least $10\%$ of the
time to warm up the system. 
The measurement frequency was between one measurement per 
80 MC `sweeps'\footnote{A sweep for the chain Metropolis algorithm
is defined as $N$ trials to move chains of particles \cite{JASTER1}.} 
for $\rho=0.82$ and $N=64^2$ and one
measurement per 5000 MC sweeps for $\rho=0.89$ and $N=256^2$
since the measurement is expensive compared to
the updating steps due to the calculation of
neighbours. $\rho$ is the reduced density since we have
set the disk diameter equal to one in the whole paper. 
The number of measurement sweeps for all performed simulations is 
collected in Tab.\ \ref{table_sweeps}. 
The  measured  observables will be discussed in the following. 
%%%%%%%%%%%%%%%%%%%%%%%%%%%%%%%%%%%%%%%%%%%%%%%%%%%%%%%%%%%%%%%%%%%%%%%%
\begin{table}[t] 
\begin{center}
\parbox{14.5cm}{\caption{ \label{table_sweeps}
Number of measurement sweeps that were performed with 
the chain Metropolis algorithm. The acceptance rate was 
between $50\%$ and $70\%$. `8 $\times$ 1500' denotes 8 independent
data sets with $1.5 \times 10^6 $ sweeps.
}}
\end{center}
\begin{center}
\begin{tabular}{lrrrr}
\vspace*{-4.0mm} \\
\hline
\hline
\multicolumn{1}{c}{$\rho$} &
\multicolumn{4}{c}{sweeps$/10^3$} \\
 & $N=32^2$ & $N=64^2$ & $N=128^2$ &  $N=256^2$  \\
\hline
0.820 &     & 7 $\times$ 1150 &     &     \\ 
0.830 &     & 7 $\times$ 1200 &     &     \\
0.840 &     & 7 $\times$ 1040 &     &     \\
0.850 &     & 7 $\times$ 1100 &     &     \\
0.855 &     & 7 $\times$ 840 &     &     \\
0.860 & 8 $\times$ 1500 & 7 $\times$ 960 &     &     \\ 
0.865 & 8 $\times$ 1500 & 7 $\times$ 640 & 8 $\times$ 430 &   \\
0.870 & 8 $\times$ 1600 & 7 $\times$ 620 & 8 $\times$ 470 &   \\
0.875 & 8 $\times$ 1700 & 8 $\times$ 500 & 8 $\times$ 430 &   \\
0.880 & 8 $\times$ 1500 & 8 $\times$ 600 & 8 $\times$ 450 & 1900  \\ 
0.885 & 8 $\times$ 1800 & 8 $\times$ 1200 & 8 $\times$ 520 & 1900 \\
0.890 & 8 $\times$ 1800 & 8 $\times$ 1100 & 8 $\times$ 550 & 6 $\times$ 1900 \\
0.895 & 8 $\times$ 1800 & 8 $\times$ 800 & 8 $\times$ 630 &   \\
0.897 & 8 $\times$ 1800 & 8 $\times$ 1500 & 8 $\times$ 650 &   \\ 
0.898 & 8 $\times$ 1500 & 8 $\times$ 480 & 8 $\times$ 480 & 6 $\times$ 710 \\
0.900 & 8 $\times$ 1500 & 8 $\times$ 640 & 8 $\times$ 370 &     \\
0.905 & 8 $\times$ 1500 & 8 $\times$ 590 & 8 $\times$ 410 &     \\
0.910 & 15000 & 16500  &  8 $\times$ 2200 &  \\
\hline
\hline
\end{tabular}
\end{center}
\end{table}
%%%%%%%%%%%%%%%%%%%%%%%%%%%%%%%%%%%%%%%%%%%%%%%%%%%%%%%%%%%%%%%%%%%%%%%%        

\subsection*{Bond orientational order parameter and susceptibility}
The orientational order of the two-dimensional hard disk system
can be described by the (global) bond orientational order parameter 
$\psi_6$. The local value of $\psi_6$
for a particle $i$ located at $\bfx_i=(x,y)$ is given by
\be
\psi_{6,i}=
\frac{1}{N_i} \sum_j \exp \left (6\, {\mathrm{i}} \, \theta_{ij} \right ) \ ,
\en
where the sum on $j$ is over the $N_i$ neighbours of this particle
and $\theta_{ij}$ is the angle  between the particles $i$
and $j$  and an arbitrary but fixed reference 
axis. Neighbours are obtained by the Voronoi 
construction \cite{VORONOI}.  The (global) bond orientational order
parameter is then defined as
\be
\psi_6= \left | \frac{1}{N} \sum_{i=1}^N \psi_{6,i} \, \right | \ . 
\en
We measured the second and fourth moment of $\psi_6$, where the former
is related to
the susceptibility by\footnote{This 
definition yields a factor $1-2/\pi$ in the thermodynamic
limit compared to 
$\chi_6=N (\langle {\psi_6}^2 \rangle -\langle \psi_6 \rangle ^2 )$.} 
\be
\chi_6=N \langle {\psi_6}^2 \rangle.
\en
%%%%%%%%%%%%%%%%%%%%%%%%%%%%%%%%%%%%%%%%%%%%% 

\subsection*{Bond orientational correlation length}
The bond orientational correlation function is defined as
\be
g_6(\bfx - \bfx \,') = \frac{\langle \psi_6^*(\bfx \,) \,  
\psi_6(\bfx \,') \rangle}{\langle 
\rho(\bfx \,) \, \rho(\bfx \,') \rangle} \ ,
\en
where 
\be
\psi_6(\bfx \,) = \sum_{i=1}^N \delta ( \bfx - \bfx_i) \, \psi_{6,i}
\en
denotes the microscopic bond orientational  order-parameter density and
\be
\rho(\bfx \,) = \sum_{i=1}^N \delta ( \bfx - \bfx_i)
\en
is the microscopic particle density.
In the isotropic phase the bond orientational correlation
length $\xi_6$ was extracted from
the `zero-momentum' bond orientational correlation function
$g_6(x)$. This is defined as 
\be
g_6(x-x') = \frac{1}{L} \int \!\! \int dy \, dy' \, g_6(\bfx-\bfx \,') \ ,  
\en
where $L$ denotes the box length. 

\begin{figure}  
\begin{center}
\mbox{\epsfxsize=7.5cm
\epsfbox{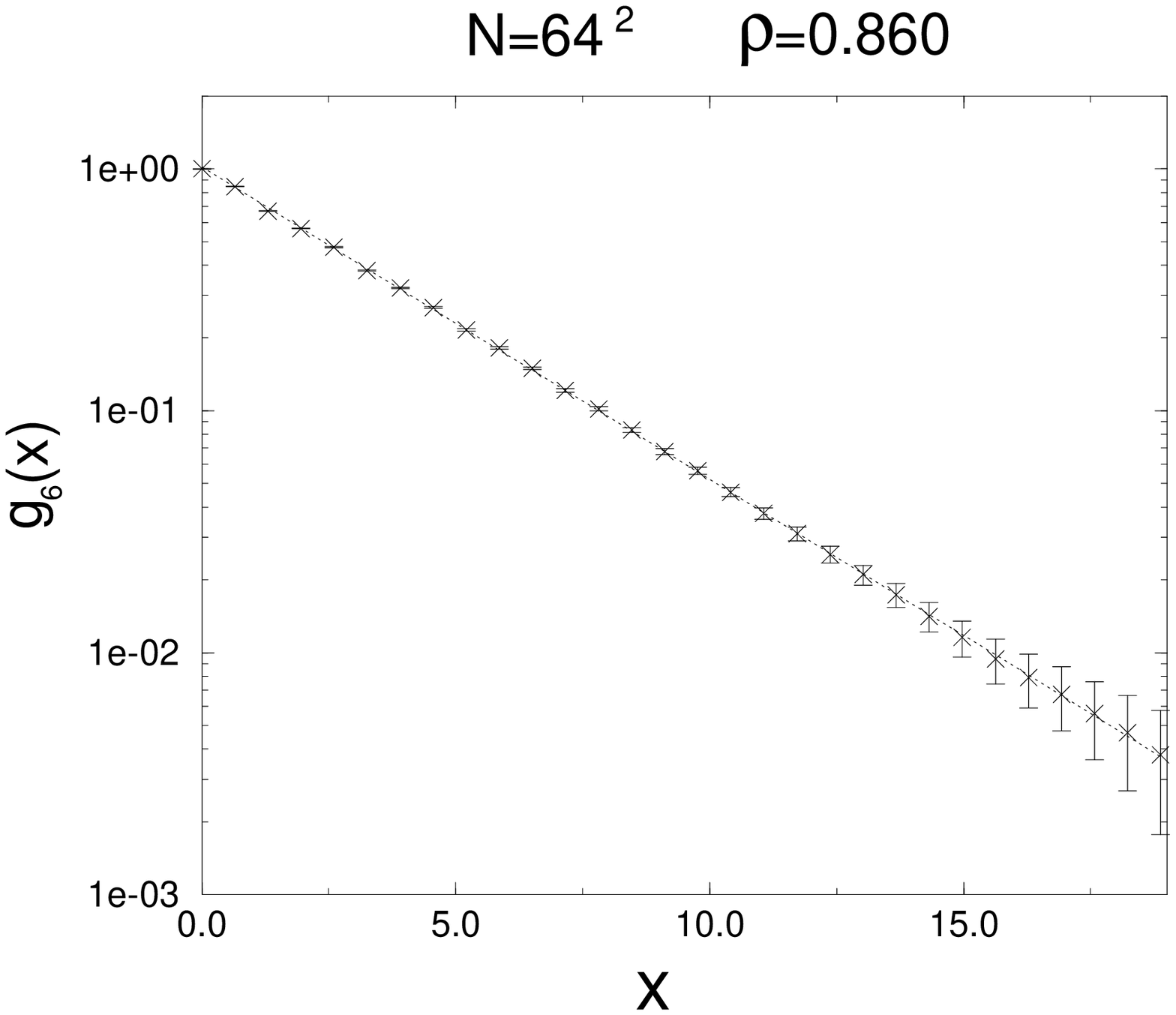}}
\hspace{2.0eM}
\mbox{\epsfxsize=7.5cm
\epsfbox{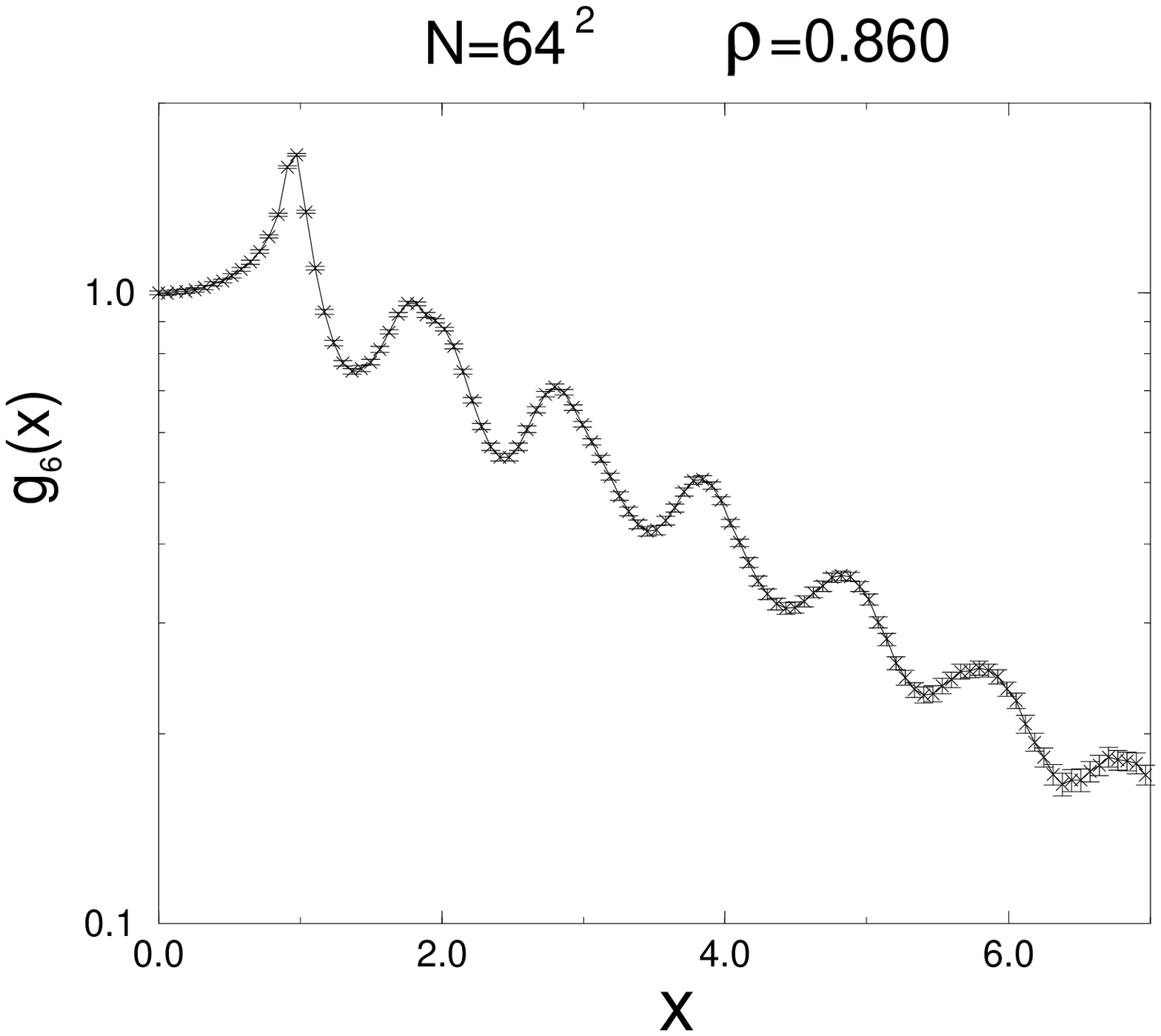}}
\parbox{14.5cm}{\caption{\label{fig_corlog}
`Zero-momentum' bond orientational correlation function $g_6(x)$ for
$N=64^2$ particles at $\rho=0.86$ (with arbitrary chosen nomalization). 
The left figure shows the
exponential behaviour for large distances, where $\Delta x$ 
was about $0.65$. The dotted line is the best fit with a hyperbolic
cosine ansatz.
The  right figure illustrates oscillations in $g_6(x)$ for small
distances (with $\Delta x \approx 0.065$). The line is a guide to the 
eye.
}}
\end{center}
\end{figure}

\begin{figure}  
\begin{center}
\centerline{\epsfxsize=9.0cm
\epsfbox{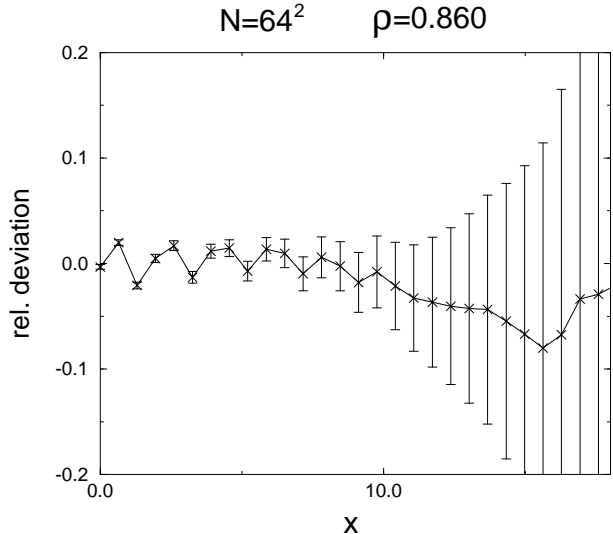}}
\parbox{14.5cm}{\caption{\label{fig_correl}
Relative deviations of $g_6(x)$ (left picture in Fig.\ 1) 
from a hyperbolic cosine fit.
}}
\end{center}
\end{figure}

In practice we use the definition
\begin{equation}
g_6(x) \sim \left \langle  \left (
\frac{1}{N_k} \int_0^L \! dy  \int_{x-\Delta x/2}^{x+\Delta x/2} \! dx \, 
\psi_6(\bfx \,) 
\right ) ^* \, \left (
\frac{1}{N_k'} \int_0^L \! dy'  \int_{-\Delta x /2}^{\Delta x/2} \! dx' \, 
\psi_6(\bfx \,')
\right ) \right \rangle \ ,
\end{equation}
where 
\be
N_k= \int_0^L \! dy  \int_{x-\Delta x/2}^{x+\Delta x/2} \! dx \, \rho(\bfx \,) \ ,
\en 
\be
N_k'= \int_0^L \! dy  \int_{-\Delta x/2}^{\Delta x/2} \! dx \, \rho(\bfx \,) \ .
\en 
Therefore, the distance 
between two particles in  $x$-direction  is not exactly
$x$, but lies between $x- \Delta x$ and $x+ \Delta x$. 
Nevertheless, assuming a pure exponential behaviour of the 
correlation function $g_6(x)$ the integration over $\Delta x$ causes no 
error. In the simulations 
the value of $\Delta x$ was given by the length of a cell of the cell
structure, i.e.\ $\Delta x \approx 2/3$, where the exact value depends
on $\rho$ and $N$.  $g_6(x)$
was fitted with a $\cosh((L/2-x)/\overline{\xi}_6)$ in the
interval $x_{\mathrm{min}} \leq x \leq L/2 $, where $\overline{\xi}_6$ and
$\xi_6$ are related by
\be
\frac{1}{2 \xi_6}=\sinh \left ( \frac{1}{2 \overline{\xi}_6} \right ) \ .
\en
To determine the influence of `excitations' we compare the results
for different minimal distances $x_{\mathrm{min}}$. The correlations 
are always dominated by the lowest state of the transfer
operator `Hamiltonian', so that it was not necessary to omit points.
In Fig.\ \ref{fig_corlog} we plot the correlation function for the
$N=64^2$ particle system at $\rho=0.86$ (with arbitrary normalization). 
The left figure shows the
correlation function as obtained from the simulation, i.e.\ 
with $\Delta x \approx 2/3$. The curve shown is the best fit
with a $\cosh$-like behaviour. As one can see there are no 
influences of `excitations'. Although the fit seems to
be consistent with the data,
there are large deviations. The reason is an oscillating 
behaviour of $g_6(x)$ as shown in the right figure, where
we have chosen $\Delta x$ ten times smaller. The same oscillations
can be seen, if we plot the relative deviations between the data
of the first case ($\Delta x \approx 2/3$) and the hyperbolic cosine fit.
This is done in Fig.\ \ref{fig_correl}.
Since the oscillating length is about $1$, the curve can be smoothed 
if one chooses $\Delta x \approx 1$. Nevertheless,
also with $\Delta x \approx 2/3$ a precise determination of 
the correlation length is possible.
Systematic errors coming from the oscillations are taken into
account. These errors get dominant
 --- compared to our statistical errors ---
for small values of $\xi_6$.
%%%%%%%%%%%%%%%%%%%%%%%%%%%%%%%%%%%%%%%%%%%%% 
%%%%%%%%%%%%%%%%%%%%%%%%%%%%%%%%%%%%%%%%%%%%%%%%%%%%%%%%%%%%%%%%%%%%%%%%
%%%%%%%%%%%%%%%%%%%%%%%%%%%%%%%%%%%%%%%%%%%%%%%%%%%%%%%%%%%%%%%%%%%%%%%%
%%%%%%%%%%%%%%%%%%%%%%%%%%%%%%%%%%%%%%%%%%%%%%%%%%%%%%%%%%%%%%%%%%%%%%%%        
%%%%%%%%%%%%%%%%%%%%%%%%%%%%%%%%%%%%%%%%%%%%%%%%%%%%%%%%%%%%%%%%%%%%%%%%        

\subsection*{Radial bond orientational correlation function}
In the isotropic phase $g_6(\bfx \,)$ is independent of the angle. Therefore,
we use the angle averaged quantity
\be
g_6(r) \sim \langle {\psi_6}^*(0) \, \psi_6(r)
\rangle / g(r) 
\en
for an additional
calculation of $\xi_6$, 
where $g(r)$ is the (radial) pair correlation function.
The radial bond 
orientational correlation function $g_6(r)$ was fitted for large 
distances  with an ansatz of the form 
\be
\label{firg6r}
g_6(r) \sim r^{-\eta_6} \exp(-r/\xi_6) \ . 
\en

\begin{figure}  
\begin{center}
\mbox{\epsfxsize=7.5cm
\epsfbox{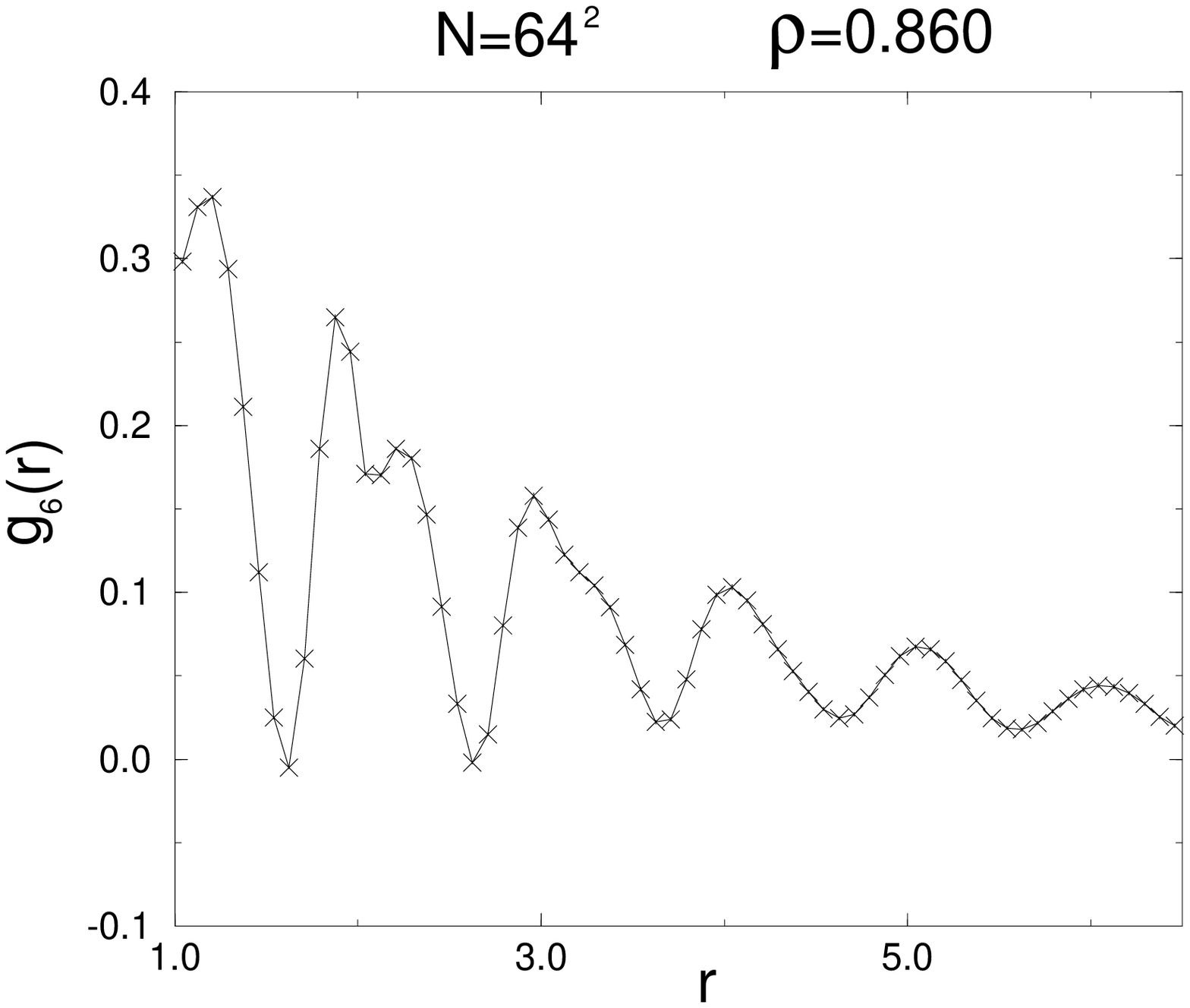}}
\hspace{2.0eM}
\mbox{\epsfxsize=7.5cm
\epsfbox{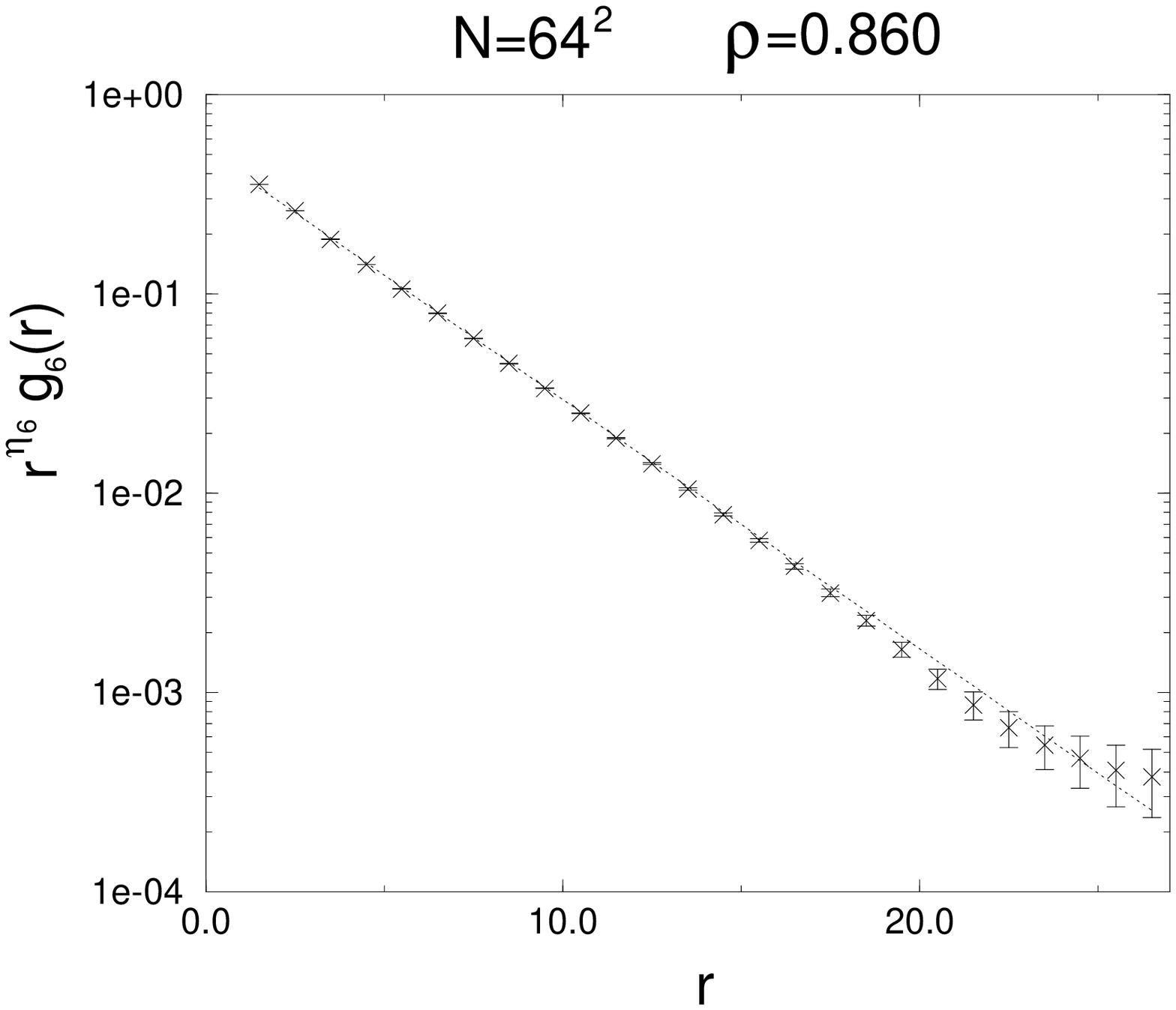}}
\parbox{14.5cm}{\caption{\label{fig_corrad}
Radial bond orientational correlation function $g_6(r)$.
The left figure shows the behaviour for small distances.
The line is just a guide to the eye.
In the right figure we plot $ r^{\eta_6}g_6(r)$ together
with an exponential fit (dotted line). 
}}
\end{center}
\end{figure}

In Fig.\ \ref{fig_corrad} we plot $g_6(r)$ for  $N=64^2$
hard disks at  $\rho=0.86$. The left figure shows 
the oscillating behaviour of $g_6(r)$. In order to smooth
the curves, $g_6(r)$ has been averaged over a distance of one.
This was done in the right figure, where $g_6(r)$ was additionally 
multiplied by $r^{\eta_6}$ in order to compare the data with an
exponential behaviour.
%%%%%%%%%%%%%%%%%%%%%%%%%%%%%%%%%%%%%%%%%%%%%%%%%%%%%%%%%%%%%%%%%%%%%%%%
%%%%%%%%%%%%%%%%%%%%%%%%%%%%%%%%%%%%%%%%%%%%%%%%%%%%%%%%%%%%%%%%%%%%%%%%

The values of $\xi_6$ obtained from $g_6(r)$ are affected by larger
systematic errors (compared to the previous method). The reason is
that one has to leave out the points with very small  and very  
large distances. The first points have to be omitted since
the ansatz is not valid in this case, while points with $r\approx L$
are affected by finite-size effects. In contrast to $g_6(x)$,
where the periodicy just leads to a 
$\cosh$ behaviour, $g_6(r)$ has no simple periodic behaviour.  
Therefore, one has to omit the points with large $r$.
Nevertheless, we use the radial bond orientational correlation function
for a determination of the correlation length. 
In all cases both values of $\xi_6$ coincide within the
statistical errors.
%%%%%%%%%%%%%%%%%%%%%%%%%%%%%%%%%%%%%%%%%%%%% 

\subsection*{Pressure}
The pressure was calculated from the
pair correlation function $g(r)$ in the range
$1.0 < r < 1.2$. From  200 bins we extracted the contact
value of the pair correlation function by fitting the data
with a power series of sixth order
and extrapolating to $g(1)$. The virial theorem relates 
this value to the pressure by \cite{METRO}
\be
\frac{p A_0}{NkT} =
\frac{\sqrt{3}}{2} \rho \left ( 1+ \frac{\pi}{2}\rho \, g(1) \right )\ ,
\en
where $A_0$ is the closed-packed area of the system, i.e.\ $A_0=N \sqrt{3}/2$.
Statistical errors were calculated by independent
data sets and by performing fits on the whole data sets to a Gaussian
distribution of $g(r)$ with variance $\Delta g(r)$.   
Systematic errors were estimated by changing the order of
the power series from six to five.

\begin{figure} 
\begin{center}
\centerline{\epsfxsize=9.0cm
\epsfbox{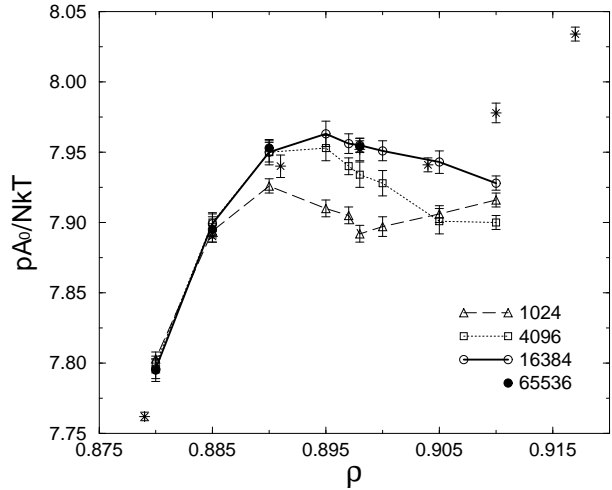}}
\parbox{14.5cm}{\caption{\label{fig_press}
Pressure as a function of the density for various system sizes.
Data of  Ref.\ [12] are marked by stars.
}}
\end{center}
\end{figure}

Our results for the pressure as a function of the system size
and the density are collected in Tab.\ \ref{table_press}
and visualized in Fig.\ \ref{fig_press}.
The quoted error is the sum of the statistical and systematic error.
The data show the end of the liquid region and the beginning
of a possible liquid-solid tie line, while no simulations in the solid phase were
made. For densities up to $\rho=0.885$, the pressure does not have any
finite-size effect within the statistical errors. Taking the finite-size
dependency of the pressure together with the data of $\xi_6$ 
(which are discussed in Sec.\ III), we find
that we have reached the thermodynamic limit 
for the systems with  $N=128^2$ particles up to $\rho=0.885$ and 
for the systems with  $N=256^2$ particles  up to
$\rho=0.89$. For densities $\rho > 0.89$
there might be still finite-size effects.
 The results are consistent with those of Zollweg and Chester 
\cite{ZOLCHE}, who used the same methods but  a rectangular box
with ratio $\sqrt{3}:2$. Only the value at $\rho=0.910$ shows
deviations. This could be a result of the square box, which
leads to larger finite size effects if the density of
the system is near the solid phase.
Another possibility is that the large systems at higher densities 
are not fully equilibrated. However, this seems to be unlikely due to
our observation of the pressure as a function of time.
Our results give a lower bound for the beginning of a coexisting
phase of $\rho \approx 0.89$, but give neither any conclusive
evidence for a first-order phase transition or an hexatic phase.
It just shows that the compressibility in this region is very high.
%%%%%%%%%%%%%%%%%%%%%%%%%%%%%%%%%%%%%%%%%%%%% 
%%%%%%%%%%%%%%%%%%%%%%%%%%%%%%%%%%%%%%%%%%%%%%%%%%%%%%%%%%%%%%%%%%%%%%%%
\begin{table}  
\begin{center}
\parbox{14.5cm}{\caption{ \label{table_press}
Pressure for densities in or near the transition region.
}}
\end{center}
\begin{center}
\begin{tabular}{lr@{.}lr@{.}lr@{.}lr@{.}l}
\vspace*{-4.0mm} \\
\hline
\hline
\multicolumn{1}{c}{$\rho$} & \multicolumn{8}{c}{$pA_0/NkT$} \\
\multicolumn{1}{c}{} & 
\multicolumn{2}{c}{$N=32^2$} &\multicolumn{2}{c}{$N=64^2$} &
\multicolumn{2}{c}{$N=128^2$}&\multicolumn{2}{c}{$N=256^2$} \\
\hline
0.880 &     7&803(5)  &  7&799(6)  &  7&796(7)  &  7&795(8)  \\ 
0.885 &     7&894(5)  &  7&900(6)  &  7&899(8)  &  7&895(9)  \\
0.890 &     7&926(5)  &  7&950(7)  &  7&950(9)  &  7&953(5)  \\
0.895 &     7&910(6)  &  7&953(9)  &  7&963(9)  &\multicolumn{2}{c}{}  \\
0.897 &     7&905(6)  &  7&940(6)  &  7&956(7)  &\multicolumn{2}{c}{}  \\ 
0.898 &     7&892(6)  &  7&934(9)  &  7&955(5)  &  7&954(4)  \\
0.900 &     7&897(7)  &  7&928(9)  &  7&951(7)  &\multicolumn{2}{c}{}  \\
0.905 &     7&906(6)  &  7&901(9)  &  7&943(8)  &\multicolumn{2}{c}{}  \\
0.910 &     7&916(5)  &  7&900(5)  &  7&928(5)  &\multicolumn{2}{c}{}  \\
\hline
\hline
\end{tabular}
\end{center}
\end{table}
%%%%%%%%%%%%%%%%%%%%%%%%%%%%%%%%%%%%%%%%%%%%%%%%%%%%%%%%%%%%%%%%%%%%%%%%        
%%%%%%%%%%%%%%%%%%%%%%%%%%%%%%%%%%%%%%%%%%%%%%%%%%%%%%%%%%%%%%%%%%%%%%%%        
%%%%%%%%%%%%%%%%%%%%%%%%%%%%%%%%%%%%%%%%%%%%%%%%%%%%%%%%%%%%%%%%%%%%%%%%

\subsection*{Local bond orientational order parameter}
\begin{figure}  
\begin{center}
\mbox{\epsfxsize=7.5cm
\epsfbox{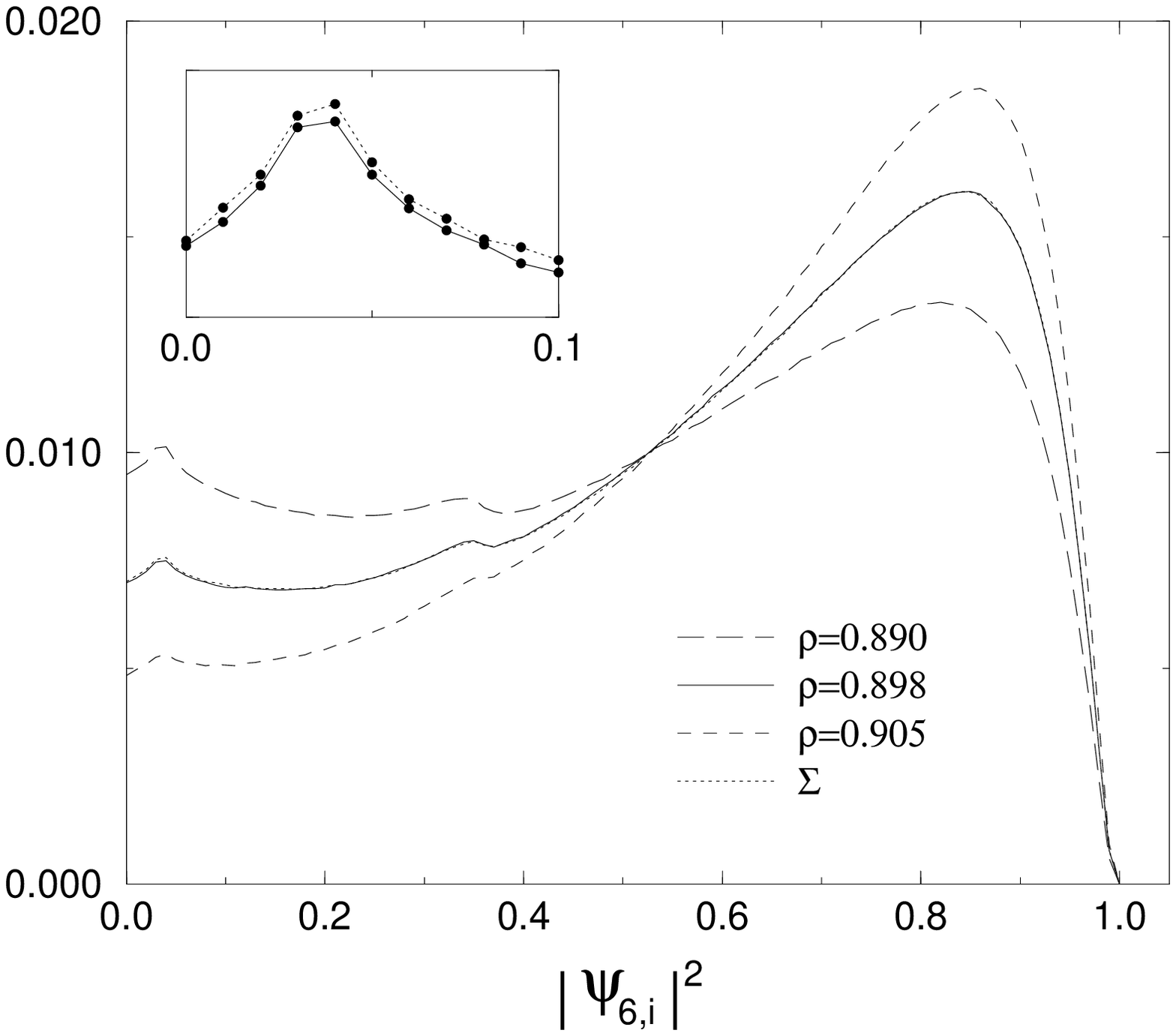}}
\hspace{2.0eM}
\mbox{\epsfxsize=7.5cm
\epsfbox{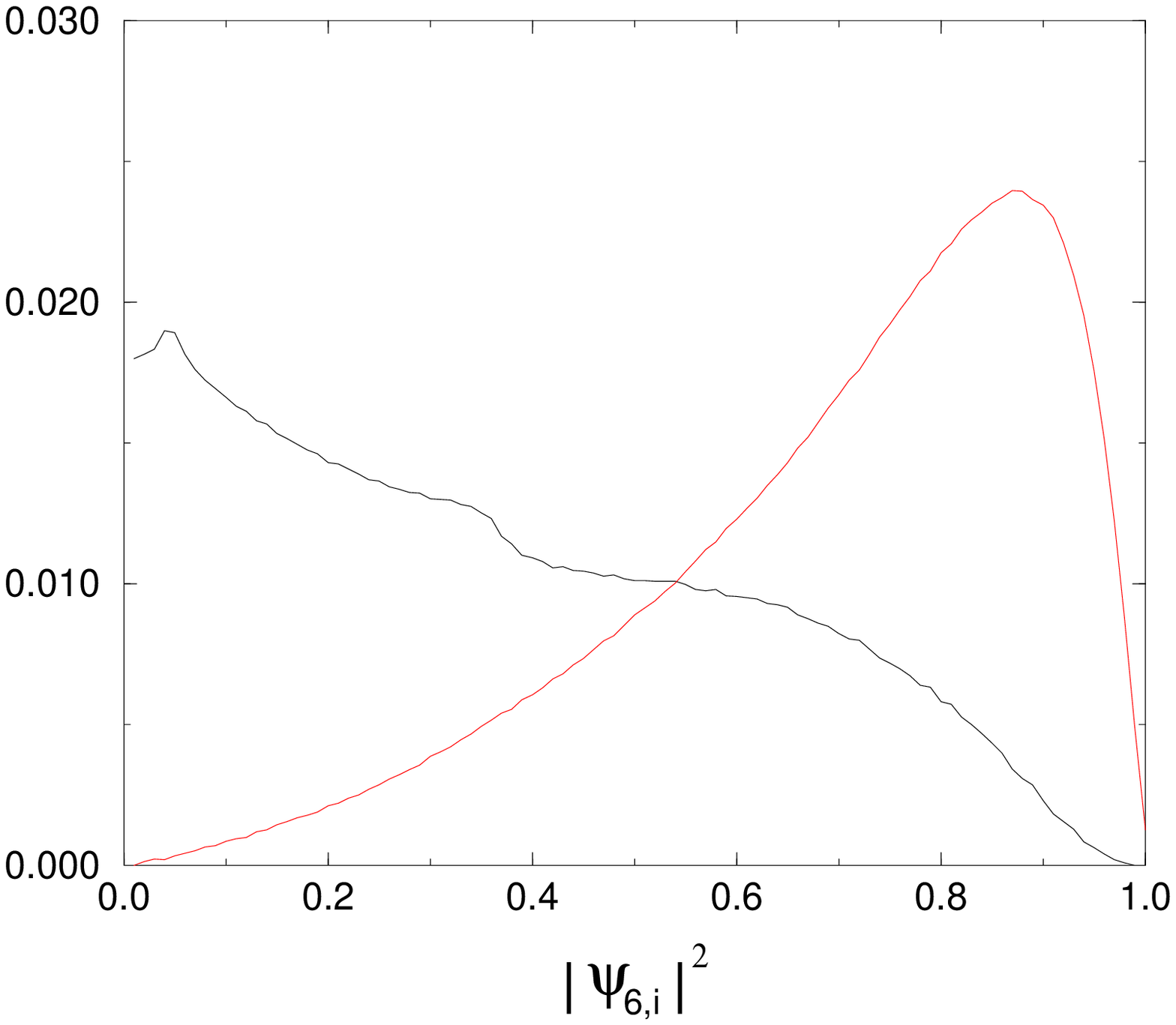}}
\parbox{14.5cm}{\caption{\label{fig_psiloksum}
The left figure shows the 
distribution occurrence of the second moment of the 
local bond orientational order parameter $\psilok$
(in arbitrary units) for $N=128^2$ hard disks 
at three different densities. $\Sigma$ indicates the curve, which is
the linear combination of $\rho_1=0.89$ and $\rho_2=0.905$.
The small inset amplifies the region with small $\psilok$
to show the small difference between the direct measurement and
the linear combination. Errors are of the order of the distance between
the two curves. The right figure displays two distributions,
which can be taken as the initial distributions for the modelling 
of all others in the transition region.
}}
\end{center}
\end{figure}

The distribution of the second moment of the local 
bond orientational order parameter $\psilok$
was first studied by Strandburg, Zollweg and Chester \cite{STZOCH}.
In the case of a first-order phase transition
(with thin interfaces) one expects that 
the distribution of the coexistence phase is the sum of 
the fluid, solid and interface 
distribution weighted with their relative areas.
On the left picture of 
Fig.\ \ref{fig_psiloksum} we plot $\psilok$ for systems
with $16384$ hard disks at three different $\rho$'s. 
To check if the distribution at $\rho=0.898$ corresponds to  a
coexisting phase, we compare it with
a combination of two other distributions at $\rho_1$ 
and $\rho_2$, respectively. It is not necessary 
that the  two chosen densities ($\rho_1$ and $\rho_2$)
are the exact values of lowest and highest density of a possible
coexisting region. This should work for two arbitrary  densities, 
provided that these systems are in the coexisting phase.
If a first-order phase transitions exists, but
the chosen  density of $\rho_1=0.89$ is too low or  $\rho_2=0.905$
is  too  high, there are deviations.
Obviously, the direct measurement and the modelling are in perfect 
agreement. Moreover, the weights of the two distributions 
correspond to their theoretical values of $8/15$ and $7/15$,
respectively. Nevertheless, an interpretation  as the sum of 
two distributions from two different phases of a first-order phase transition
makes only sense if the system size is  larger than the two interfaces.
But the results of the following sections will show that a first-order
phase transition with such small interfaces can be ruled out.
Therefore the situation is more complicate than in this simple picture.
%%%%%%%%%%%%%%%%%%%%%%%%%%%%%%%%%%%%%%%%%%%%%%%%%%%%%%%%%%%%%%%%%%%%%%%%        
%%%%%%%%%%%%%%%%%%%%%%%%%%%%%%%%%%%%%%%%%%%%%%%%%%%%%%%%%%%%%%%%%%%%%%%%        

The distributions of $\psilok$ in the transition region
can be modelled as the sum of two initial distributions.
The reconstruction of these distributions is not unique.
A decomposition
is shown in the right picture of Fig.\ \ref{fig_psiloksum}.
One of the distributions results primary from particles with six neighbours,
while the other  is mainly the
sum of distributions from particles with coordination numbers
unequal six.

Additional investigations for the dependency of the distribution of 
$\psilok$ on the system size are discussed 
in Sec.\ IV.
%%%%%%%%%%%%%%%%%%%%%%%%%%%%%%%%%%%%%%%%%%%%% 

\subsection*{Topological defects}
An analysis of the numbers of neighbours of each particle
as obtained from the Voronoi construction gives a
characterization of the defect structure of a two-dimensional system.
The average number of neighbours is --- independent of the  
state of disorder --- six.  In a perfect solid each particle has
six neighbours. Particles with any other number of neighbours 
represent a disclination. Dislocations are pairs of disclinations.
We define the density of defects as
\be
\ndef=\frac{1}{N} \sum_{i\ne 6} N_i^{\mathrm{Nb}} \ ,
\en
where $ N_i^{\mathrm{Nb}}$ denotes the number of particles with
$i$ neighbours. Alternatively one can take the strengths of
the disclinations into account and define the density of defects 
as
\be
n_{\mathrm{def}}'=
\frac{1}{N} 
\sum_i |i-6| \, N_i^{\mathrm{Nb}} 
= \frac{2}{N} \sum_{i < 6} (6-i) \, N_i^{\mathrm{Nb}} \  .
\en
However, the difference between the two definitions for 
$\rho \ge 0.88 $ is lower $1\%$. 

%%%%%%%%%%%%%%%%%%%%%%%%%%%%%%%%%%%%%%%%%%%%%%%%%%%%%%%%%%%%%%%%%%%%%%%%        
\begin{figure}[t] 
\begin{center}
\centerline{\epsfxsize=9.0cm
\epsfbox{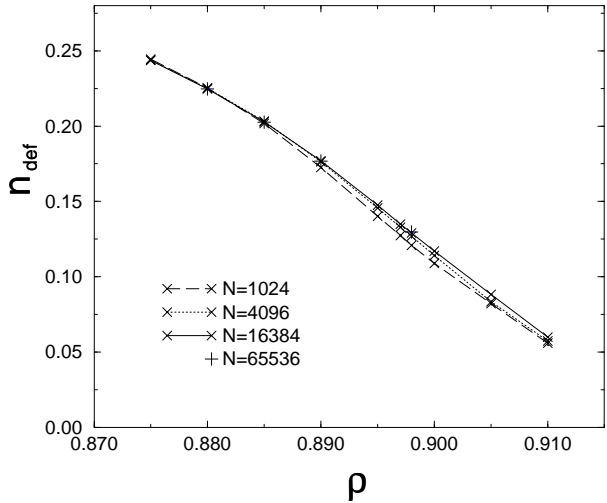}}
\parbox{14.5cm}{\caption{\label{fig_ndef}
Topological defect density $\ndef$ as a function of $\rho$.
Statistical errors are too small for a visualization.
}}
\end{center}
\end{figure}
%%%%%%%%%%%%%%%%%%%%%%%%%%%%%%%%%%%%%%%%%%%%%%%%%%%%%%%%%%%%%%%%%%%%%%%%        
In Fig.\ \ref{fig_ndef} we plot $\ndef$ as a function of $\rho$.
One can see that there is a linear behaviour of $\ndef$ for 
$\rho \ge 0.89$\footnote{A linear behaviour of a local order parameter 
was also found in Ref.\ \cite{MIWEMA}.}. 
As in the case of the distribution of  $\psilok$,
this could be explained with the coexistence of two different phases.
However, if one examines the defect structure of several configurations
in the transition region and the configurations itself, 
one finds no hint for two coexisting phases, while they are
compatible with the picture of a homogeneous phase.
Therefore, the conventional picture
(of a first-order phase transition with thin interfaces) 
of two separated  phases is incompatible with the data.
The results of Sec.\ IV will confirm this assumption.

There are different possibilities to explain this linear behaviour.
On the one hand there could be a weak first-order phase transition
with an interface width which is larger than the box length $L$ of
our largest system
of $N=256^2$ particles, on the other hand  a continuous
transition with a homogeneous phase.  In both cases increasing the
density $\rho$ primary leads to a decrease of the defect density,
since the average density in the ordered regions in higher than
the density in  unordered  regions 
(i.e.\ a disclination or dislocation needs more space than 
perfect crystalline structures).
In the case of a first-order phase transition the defects will
form some larger structure, while there is a homogenous distribution
for the continuous transition.
%%%%%%%%%%%%%%%%%%%%%%%%%%%%%%%%%%%%%%%%%%%%% 

%%%%%%%%%%%%%%%%%%%%%%%%%%%%%%%%%%%%%%%%%%%%%%%%%%%%%%%%%%%%%%%%%%%%%%%%   
%%%%%%%%%%%%%%%%%%%%%%%%%%%%%%%%%%%%%%%%%%%%%%%%%%%%%%%%%%%%%%%%%%%%%%%%%
\section{Simulation in the isotropic phase}
In the isotropic phase we measured the susceptibility 
and the correlation length of the bond
orientation. Subsequently, we compare the results with the 
predictions of the KTHNY theory, i.e.\  a
critical exponent $\eta_6$ of $1/4$ and
an exponential singularity for
the correlation length 
\be
\label{KTxi}
\xi_6(t)=a_{\xi} \, \exp \left ( b_{\xi}\, t^{-1/2}  \right ) 
\en
and the susceptibility  
\be
\label{KTchi}
\chi_6(t)=a_{\chi} \, \exp \left ( b_{\chi}\, t^{-1/2} \right )
\en
if $t=\rhoi -\rho \rightarrow 0^+$.
A detailed description of these measurements is given
in Ref.\ \cite{JASTER1}.
%%%%%%%%%%%%%%%%%%%%%%%%%%%%%%%%%%%%%%%%%%%%%%%%%%%%%%%%%%%%%%%%%%%%%%%%
\begin{table}[t] 
\begin{center}
\parbox{14.5cm}{\caption{ \label{table_xichi}
Bond orientational correlation length $\xi_6$ and
susceptibility $\chi_6$ for various densities
in the isotropic phase. $N$ refers to the system
sizes used.
}}
\end{center}
\begin{center}
\begin{tabular}{lcr@{.}lr@{.}l}
\vspace*{-4.0mm} \\
\hline
\hline
\multicolumn{1}{c}{$\rho$} &\multicolumn{1}{c}{$N$} &
\multicolumn{2}{c}{$\xi_6$} &
\multicolumn{2}{c}{$\chi_6$} \\
\hline
0.82   &  $64^2$  &            1&513(50)  &   3&797(13) \\
0.83   &  $64^2$  &            1&800(35)  &   4&693(15) \\
0.84   &  $64^2$  &            2&156(40)  &   6&052(24) \\   
0.85   &  $64^2$  &            2&635(30)  &   8&415(41) \\  
0.855  &  $64^2$  &            2&995(35)  &  10&30(6)   \\  
0.86   &  $64^2$  &            3&425(40)  &  12&96(9)   \\  
0.865  &  $64^2$, $128^2$ &    4&14(10)   &  17&45(18)  \\ 
0.87   &  $128^2$ &            5&03(15)   &  25&00(39)  \\ 
0.875  &  $128^2$ &            6&65(30)   &  39&5(8)    \\ 
0.88   &  $128^2$, $256^2$ &   9&56(26)   &  75&0(21)   \\  
0.885  &  $128^2$, $256^2$ &  15&65(51)   &  176&8(61)  \\
0.89   &  $256^2$ &           38&0(15)    & \multicolumn{2}{l}{865(44)} \\ 
\hline
\hline
\end{tabular}
\end{center}
\end{table}
%%%%%%%%%%%%%%%%%%%%%%%%%%%%%%%%%%%%%%%%%%%%%%%%%%%%%%%%%%%%%%%%%%%%%%%%        

Our results of $\chi_6$ and $\xi_6$ as a function of the density 
are summarized in Tab.\ \ref{table_xichi}.
We analyzed the critical behaviour of $\chi_6(\rho)$ and $\xi_6(\rho)$
by performing least square fits according to Eqs.\
(\ref{KTxi}) and (\ref{KTchi}).
Using all 12  points   we got a
$\chi^2$ per degree of freedom (d.o.f.) of $0.75$ 
for $\xi_6(t)$ and $0.65$ for $\chi_6(t)$, i.e.\ 
the data are in a very good agreement with
an exponential singularity of the KTHNY type.
This is not  only a result of large statistical errors as  can be seen 
if one uses different approaches for the singularities.
For example,
a conventional second-order behaviour with a power-law singularity
of the form $\ln(\xi_6)= a - \nu \ln(t)$ yields $\chi^2/$d.o.f.$=4.1$.

%%%%%%%%%%%%%%%%%%%%%%%%%%%%%%%%%%%%%%%%%%%%%%%%%%%%%%%%%%%%%%%%%%%%%%%%
\begin{figure}[t]  
\begin{center}
\centerline{\epsfxsize=9.0cm
\epsfbox{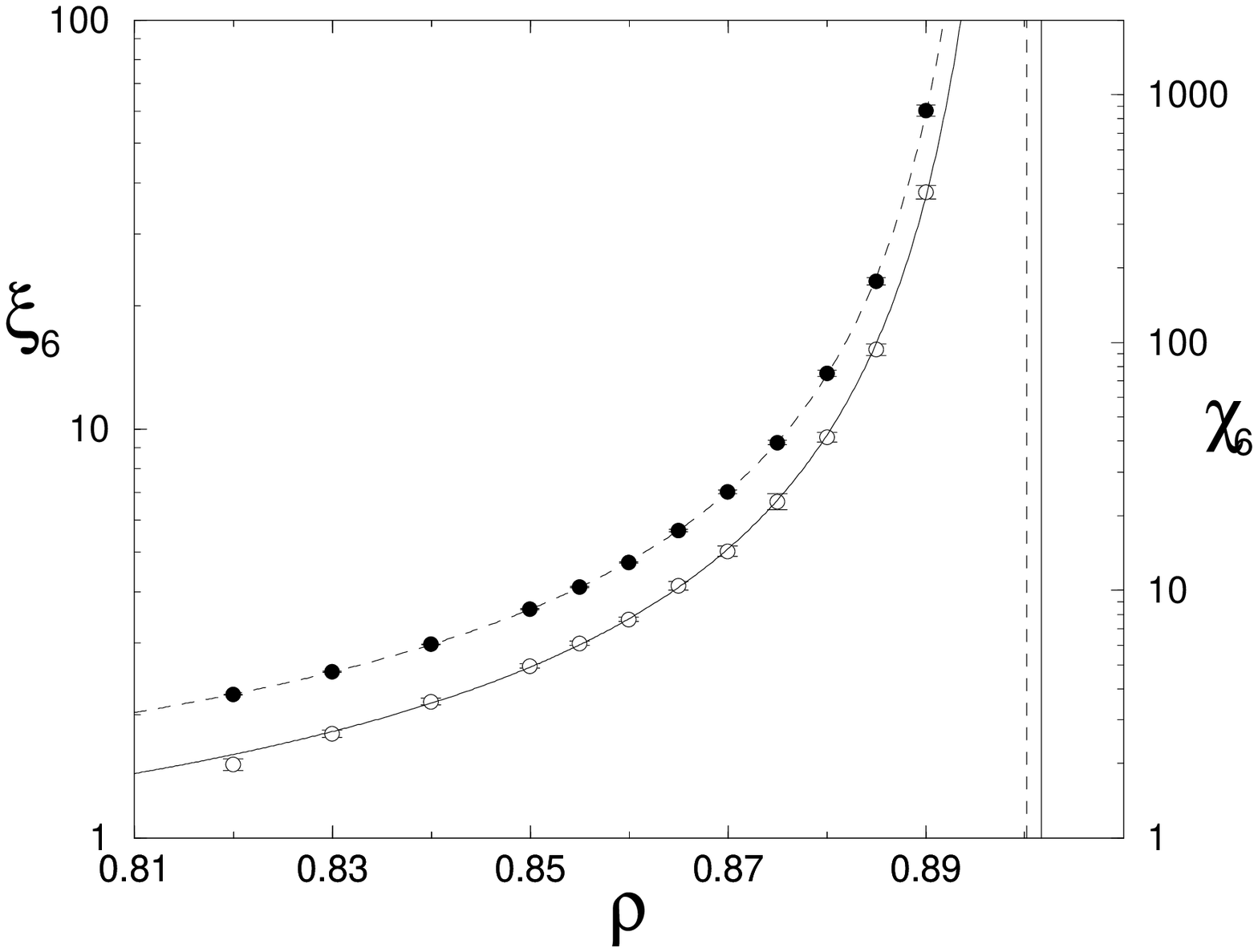}}
\parbox{14.5cm}{\caption{\label{fig_sing}
Susceptibility (full symbols) and bond orientational correlation length
(open symbols) 
as a function of density. The curves shown
are the best fits for a KTHNY behaviour 
(for all measured points). The critical values of
$\rho$ are visualized by  vertical lines.
}}
\end{center}
\end{figure}
%%%%%%%%%%%%%%%%%%%%%%%%%%%%%%%%%%%%%%%%%%%%%%%%%%%%%%%%%%%%%%%%%%%%%%%%
All results for the fit parameter  are collected in Tab.\ 
\ref{table_fit}. 
The values for $\chi_6(\rho)$ and $\xi_6(\rho)$ together with 
the fitted curves are
shown in  Fig.\ \ref{fig_sing}. 
We  also made fits where we have omitted some data
at lower densities. The fit parameters for $\chi_6(\rho)$
show only non-essential changes, while the changes for $\xi_6(\rho)$
are of the order of the statistical errors.
An analysis of the behaviour of $\ln ( \chi_6/{\xi_6}^{7/4} )$
as a function of $\ln ( \xi_6 )$ yields the following value of the 
critical exponent:
\be
\eta_6=0.251(36) \ .
\en
%%%%%%%%%%%%%%%%%%%%%%%%%%%%%%%%%%%%%%%%%%%%%%%%%%%%%%%%%%%%%%%%%%%%%%%%
\begin{table}[b]  
\begin{center}
\parbox{14.5cm}{\caption{ \label{table_fit}
Best fit parameter for the critical behaviour of the correlation 
length and susceptibility. For $0.82 \le \rho \le 0.89$ we
have fitted 12 points, while we used 8 points in the range
$0.855 \le \rho \le 0.89$.
}}
\end{center}
\begin{center}
\begin{tabular}{crr@{.}lr@{.}lr@{.}lr@{.}l}
\vspace*{-4.0mm} \\
\hline
\hline
\multicolumn{1}{c}{Fit} &\multicolumn{1}{c}{range} &
\multicolumn{2}{c}{$\ln(a)$} & \multicolumn{2}{c}{$b$} &
\multicolumn{2}{c}{$\rhoi$} & \multicolumn{2}{c}{$\chi^2$/d.o.f.} \\
\hline
$\xi_6(\rho)$  &  $0.82 \le \rho \le 0.89$  &  -1&44(8)  &  0&547(21)  
 &   0&9017(6)  &  0&75 \\
$\xi_6(\rho)$  &  $0.855 \le \rho \le 0.89$  &  -1&27(13)  &  0&505(31) 
 &   0&9006(8)  &  0&23 \\
$\chi_6(\rho)$  &  $0.82 \le \rho \le 0.89$  &   -1&65(3)  &  0&847(7) 
 &   0&9002(3)  &  0&65 \\
$\chi_6(\rho)$  &  $0.855 \le \rho \le 0.89$  &  -1&60(9)  &  0&834(21) 
 &   0&9000(4)  &  0&58 \\
\hline
\hline
\end{tabular}
\end{center}
\end{table}
%%%%%%%%%%%%%%%%%%%%%%%%%%%%%%%%%%%%%%%%%%%%%%%%%%%%%%%%%%%%%%%%%%%%%%%%        

%%%%%%%%%%%%%%%%%%%%%%%%%%%%%%%%%%%%%%%%%%%%%%%%%%%%%%%%%%%%%%%%%%%%%%%%   
%%%%%%%%%%%%%%%%%%%%%%%%%%%%%%%%%%%%%%%%%%%%%%%%%%%%%%%%%%%%%%%%%%%%%%%%%
\section{Simulation in the transition region}
We now come to the simulations with $\rho \approx \rhoi$.
Finite-size scaling (FSS) implies for the susceptibility
\be
\chi_6 \sim L^{2-\eta_6} \, f(L/\xi_6) \ .
\en
Assuming the prediction of the KTHNY theory,
the correlation length $\xi_6$ diverges  at $\rho=\rhoi$   and
$f$ is a constant independent of $L$. We use this FSS behaviour
to locate $\rhoi$, where we take $\eta_6=1/4$. 
In the hexatic phase ($\rhoi < \rho \le \rhom $)
the correlation length $\xi_6$ also diverges, so that 
$f$ is still independent of $L$. In this phase
$\eta_6$ is a decreasing function of the density
which goes to zero if $\rho$ approaches  the melting density
$\rhom$, i.e.\ at the end of the hexatic phase.
For $\rho$'s below 
$\rhoi$, one has to take corrections of $\chi_6 \sim L^{2-\eta_6}$
for finite correlations lengths
into account. Our results for the susceptibility are collected
in Tab.\ \ref{table_psi}.
%%%%%%%%%%%%%%%%%%%%%%%%%%%%%%%%%%%%%%%%%%%%%%%%%%%%%%%%%%%%%%%%%%%%%%%%
\begin{table}
\begin{center}
\parbox{14.5cm}{\caption{ \label{table_psi}
The susceptibility per particle in the transition region.
}}
\end{center}
\begin{center}
\begin{tabular}{lllll}
\vspace*{-4.0mm} \\
\hline
\hline
\multicolumn{1}{c}{$\rho$} & \multicolumn{4}{c}{$\chi_6/N$} \\
\multicolumn{1}{c}{} & 
\multicolumn{1}{c}{$N=32^2$} &\multicolumn{1}{c}{$N=64^2$} &
\multicolumn{1}{c}{$N=128^2$}&\multicolumn{1}{c}{$N=256^2$} \\
\hline
 0.895  &  0.2620(9) &   0.1970(17) &   0.1409(24) &     \\
 0.897  &  0.2987(9) &   0.2418(18) &   0.1899(25) &     \\
 0.898  &  0.3175(10) &   0.2612(17) &   0.2160(25) &   0.1788(29) \\
 0.900  &  0.3514(10) &   0.3076(13) &   0.2630(17) &     \\
 0.905  &  0.4235(19) &   0.4055(11) &   0.3745(13) &     \\
 0.910  &  0.4900(29) &   0.4840(24) &   0.4707(10) &     \\
\hline
\hline
\end{tabular}
\end{center}
\end{table}
%%%%%%%%%%%%%%%%%%%%%%%%%%%%%%%%%%%%%%%%%%%%%%%%%%%%%%%%%%%%%%%%%%%%%%%%

If we use the FSS behaviour to locate $\rhoi$ and $\rhom$, then
the requirement of $\eta_6(\rhoi)=1/4$ yields \cite{JASTER1}
\be
\rhoi=0.899(1) \ ,
\en
while  $\eta_6(\rhom)=0$ leads to the estimate $\rhom \gsim 0.91$.
The value of $\rhoi$ is in agreement with that 
obtained from the singularities of $\xi_6(t)$ and $\chi_6(t)$.
A slightly different value of $\eta_6$ (from the relation of $\chi_6$
and $\xi_6$ in the isotropic phase) would not change this situation.
Moreover, our values of $\rhoi$ are in agreement with the result
of Weber et al.\ \cite{WEMABI} obtained 
from the fourth-order cumulant intersection
($\rhoi=0.8985(5)$). However, it differs from  their
value  obtained using the singularity of $\chi_6$ ($\rhoi=0.913$).
The result $\rhoi=0.916(4)$ of Fern\'{a}ndez et al.\ \cite{FEALST} 
is not compatible with our value.

%%%%%%%%%%%%%%%%%%%%%%%%%%%%%%%%%%%%%%%%%%%%%%%%%%%%%%%%%%%%%%%%%%%%%%%%        
\begin{figure}
\begin{center}
\centerline{\epsfxsize=9.0cm
\epsfbox{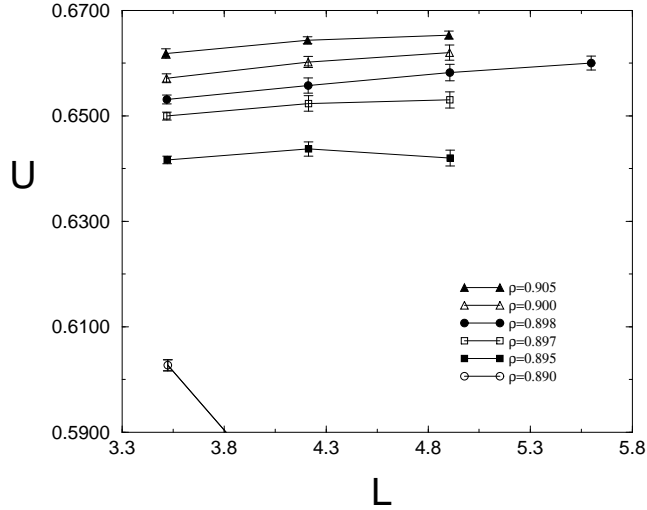}}
\parbox{14.5cm}{\caption{\label{fig_FSSU}
Finite-size scaling of the cumulant in the transition region.
}}
\end{center}
\end{figure}
%%%%%%%%%%%%%%%%%%%%%%%%%%%%%%%%%%%%%%%%%%%%%%%%%%%%%%%%%%%%%%%%%%%%%%%%        
Another quantity which can be used to analyze the kind of the transition
is the fourth-order cumulant 
\be
U=1-\frac{ \left \langle {\psi_6}^4 \right \rangle}{3 \left 
\langle {\psi_6}^2  \right \rangle ^2} \ .
\en
According to the prediction of the KTHNY theory, 
$U$ should be independent of the system size $L$
in the whole hexatic phase. 
In contrast to this, in the case of a conventional first-order
phase transition there is only a single point, where the cumulants
of different system sizes collapse. 
Since there is a large region between $\rhoi \approx 0.9$ and
$\rhom \gsim 0.91$, the behaviour of $U$ can be used to distinguish
between a KTHNY and a first-order transition.
The intersection of the cumulant $U$ in a single point was
an argument in Ref.\ \cite{WEMABI} against the existence of
an hexatic phase.  Unfortunately, statistical errors in
our data are too large to answer this question 
as can be seen in Fig.\ \ref{fig_FSSU}.

\begin{figure}
\begin{center}
\centerline{\epsfxsize=9.0cm
\epsfbox{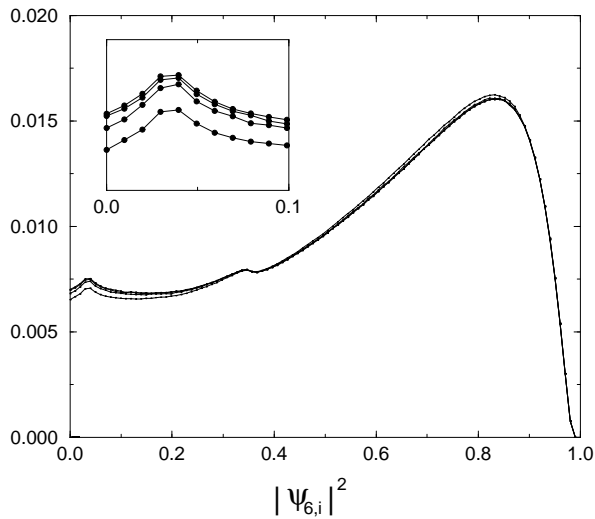}}
\parbox{14.5cm}{\caption{\label{fig_psilok}
Distribution of the second moment of the
local bond orientational order parameter (in arbitrary
units) for different system sizes. The small inset amplifies the
region with small $\psilok$, where are the largest deviations. 
Statistical errors are of the order of the symbols in the inset.
}}
\end{center}
\end{figure}

Another possibility to distinguish a first-order phase transition
from a continuous transition is to study the dependency of the
distribution of $\psilok$ on the size of the system.
If the system exhibits a homogeneous hexatic phase, then changing
the size of the system should not lead to any changes in the distribution.
On the other hand, if the transition is of first order 
one would expect that the distribution is a combination of the
solid, fluid and interface distribution. Therefore, changing the size of the
system would result in a change of  the distribution,
because the area of the interface scales only linear with $L$.
In Fig.\ \ref{fig_psilok} we plot $\psilok$ at $\rho=0.898$
for four different system sizes. Apart from finite-size effects, 
which getting weaker for larger systems, no difference
between the distributions can be seen. The distributions for
the two largest systems coincide within statistical errors.
Therefore, one can rule out a first-order transition
with  thin interfaces\footnote{The results are also compatible
with the occurrence of two very small interfaces (i.e.\ a width
of ${\cal O}(1)$), but this can be
ruled out  due to the examination of the defect structure of several
configurations.}, while a  first-order transition 
with an interface width larger than the largest system size 
$L$ and a continuous transition are compatible with the data.
These results coincide with those of Fern\'{a}ndez et al.\ 
\cite{FEALST}, who performed similar measurements in the $NpT$
ensemble using a rectangular box of ratio $\sqrt{3}:2$.
The data of Fig.\ \ref{fig_psilok}
show also that the chosen ratio of the side lengths of $1:1$
causes no large finite-size effects.
%%%%%%%%%%%%%%%%%%%%%%%%%%%%%%%%%%%%%%%%%%%%%%%%%%%%%%%%%%%%%%%%%%%%%%%%        
%%%%%%%%%%%%%%%%%%%%%%%%%%%%%%%%%%%%%%%%%%%%%%%%%%%%%%%%%%%%%%%%%%%%%%%%        

%%%%%%%%%%%%%%%%%%%%%%%%%%%%%%%%%%%%%%%%%%%%%%%%%%%%%%%%%%%%%%%%%%%%%%%%        
%%%%%%%%%%%%%%%%%%%%%%%%%%%%%%%%%%%%%%%%%%%%%%%%%%%%%%%%%%%%%%%%%%%%%%%%        
\section{Conclusions and outlook}

We presented a detailed Monte Carlo study of the two-dimensional
hard disk model in the $NVT$ ensemble. 
The investigations were performed in the
isotropic phase and in the transition region.

The behaviour of the defect density as well as the
distribution of the local order parameter
in the transition region were in good agreement with a simple model
of two coexisting phases, i.e.\ the data could be modelled as 
the sum of two different phases, where the relative areas 
of the two phases
are proportional to $\rho$. However, the defect structure
of the system and the distribution of $\psilok$ as a function of $L$
showed, that there are not two separated phases with a thin interface. 
The data can be explained by a  weak first-order transition 
with a width of the interface which is 
larger than the largest system size $L$ or by a 
continuous transition with a homogeneous phase.

The behaviour of the pressure was compatible with both
a first-order and a KTHNY-like
scenario. The data just give a lower limit of $\rho\approx 0.89$ for 
the coexisting phase.

In the isotropic phase we examined the dependency of the
correlation length and the susceptibility on the density $\rho$.
We showed that the data are in good agreement with the
prediction of an exponential singularity from 
the KTHNY theory. The critical exponent $\eta_6$
was derived from the relation on $\xi_6$ and $\chi_6$ for
$\xi_6 \rightarrow \infty$. We got $\eta_6=0.251(36)$, which coincides with 
the prediction $\eta_6=1/4$.

The simulations in the transition region were used to
measure the finite-size scaling of the susceptibility.
The value of $\rhoi=0.899(1)$ (assuming $\eta_6=1/4$)
coincided with those from the
KTHNY-like behaviour of $\xi_6(\rho)$ and $\chi_6(\rho)$. 
Furthermore, the requirement $\eta_6( \rhom )=0$ led to the
estimate $\rhom \gsim 0.91$.
The data of the fourth-order cumulant $U$ were affected by
too large statistical errors in order to draw any conclusives.

In summary, all data are compatible with a KTHNY-like phase transition.
A one-stage continuous transition ($\rhoi=\rhom$) as proposed in 
Ref.\ \cite{FEALST} and a first-order transition with small
correlation length can be ruled out\footnote{Similar results are obtained
for an $r^{-12}$ repulsive potential by Bagchi, Andersen
and Swope \cite{BAANSW}.}. 
Further numerical investigations
have to be performed to make a clear decision between a 
weak first-order phase transition and a continuous scenario.
This could be done for example by studying the positional order
in the transition region. Work along this line is in progress. 

%%%%%%%%%%%%%%%%%%%%%%%%%%%%%%%%%%%%%%%%%%%%%%%%%%%%%%%%%%%%%%%%%%%%%%%%
%%%%%%%%%%%%%%%%%%%%%%%%%%%%%%%%%%%%%%%%%%%%%%%%%%%%%%%%%%%%%%%%%%%%%%%%
\section*{Acknowledgments}
We thank Harro Hahn for helpful discussions and the 
Institute of Scientific Computing in  Braunschweig
for providing computer time on their CRAY T3E.
% Helpful comments on our draft by ..... are greatfully acknowledged.
Especially we benefit from discussions with Rainer Gensch. 
 
%%%%%%%%%%%%%%%%%%%%%%%%%%%%%%%%%%%%%%%%%%%%%%%%%%%%%%%%%%%%%%%%%%%%%%%%
%%%%%%%%%%%%%%%%%%%%%%%%%%%%%%%%%%%%%%%%%%%%%%%%%%%%%%%%%%%%%%%%%%%%%%%%        

\end{document}